\documentclass[submission,copyright,creativecommons]{eptcs}
\providecommand{\event}{UITP 2014}
\usepackage{breakurl}

\usepackage{graphicx}
\usepackage[export]{adjustbox}
\usepackage[utf8]{inputenc}
\usepackage{ifthen}
\usepackage{isabelle,isabellesym}

\isadroptag{theory}
\isabellestyle{literal}

\newcommand{\secref}[1]{\S\ref{#1}}
\newcommand{\figref}[1]{figure~\ref{#1}}
\newcommand{\Figref}[1]{Figure~\ref{#1}}

\renewcommand{\hyperlink}[2]{#2}  

\begin{document}

\title{System description: Isabelle/jEdit in 2014}
\author{Makarius Wenzel \thanks{Research supported by Project
    Paral-ITP (ANR-11-INSE-001).}
\institute{Univ. Paris-Sud, Laboratoire LRI, UMR8623, Orsay, F-91405, France \\
  CNRS, Orsay, F-91405, France}}
\def\titlerunning{Isabelle/jEdit in 2014}
\def\authorrunning{M. Wenzel}
\maketitle

\begin{abstract}
  This is an updated system description for Isabelle/jEdit, according to the
  official release Isabelle2014 (August 2014). The following new PIDE
  concepts are explained: asynchronous print functions and document
  overlays, syntactic and semantic completion, editor navigation,
  management of auxiliary files within the document-model.
\end{abstract}

\begin{isabellebody}%
\def\isabellecontext{Paper}%
\isadelimtheory
\endisadelimtheory
\isatagtheory
\isacommand{theory}\isamarkupfalse%
\ Paper\isanewline
\isakeyword{imports}\ Main\ {\isachardoublequoteopen}{\isachartilde}{\isachartilde}{\isacharslash}src{\isacharslash}Doc{\isacharslash}Isar{\isacharunderscore}Ref{\isacharslash}Base{\isachardoublequoteclose}\isanewline
\isakeyword{begin}%
\endisatagtheory
{\isafoldtheory}%
\isadelimtheory
\endisadelimtheory
\isamarkupsection{Introduction%
}
\isamarkuptrue%
\begin{isamarkuptext}%
Isabelle/jEdit is a Prover IDE (PIDE) that integrates \emph{parallel
proof checking} \cite{Wenzel:2009,Wenzel:2013:ITP} with \emph{asynchronous
user interaction}
\cite{Wenzel:2010,Wenzel:2012:UITP-EPTCS,Wenzel:2014:ITP-PIDE}, based on a
document-oriented approach of \emph{continuous proof processing}
\cite{Wenzel:2011:CICM,Wenzel:2012:CICM}. This enables the user to edit
whole libraries of formalized mathematics directly in the editor, with
real-time visualization of feedback produced by the prover. Today
Isabelle/jEdit is the default user-interface for Isabelle, but this has
required many years of developing the PIDE concepts and getting the
underlying Isabelle/Scala infrastructure into a robust and scalable state.
The ultimate goal is to load the whole \emph{Archive of Formal Proofs}
\cite{AFP} into a single IDE session, but that is growing at a high
rate,\footnote{In September 2014, AFP consisted of 196 articles in 2144
source files, comprising 51\,MB total. Checking all of that in batch mode
takes approximately 10\,h CPU time and 1\,h elapsed time on a solid 8-core
Intel Xeon workstation.} and there are still theory name space problems
preventing that.

\medskip Although Isabelle/jEdit is the most visible Prover IDE application,
and sometimes people erroneously attach the label ``jEdit'' to anything
coming after the TTY loop and Proof General \cite{Aspinall:TACAS:2000} in
Isabelle, the PIDE principles are meant to be more general and applicable to
other front-ends.

Isabelle/jEdit is an example for a \emph{rich-client application} that is
run on the local machine, with non-trivial resource requirements: 2--4 CPU
cores and 2--4 GB memory minimum. An interesting alternative is the
\emph{client-server application} Clide
\cite{Lueth-Ring:2013,Lueth-Ring:2014}, which combines Isabelle/Scala/PIDE
with recent Web technology on the JVM, and supports collaborative
interactive theorem proving in particular.

\medskip Here is a brief historical overview of Isabelle/jEdit so far:

\begin{itemize}

\item In 2005 all major CPU manufacturers started to ship multicore systems
for the consumer market. Ever since the burden of explicit parallelism has
been imposed on application developers, in order to keep up with the changed
side-conditions of \emph{Moore's Law}, and participate in continued
performance improvements of computing hardware.

\item In 2006--2008 Isabelle and its underlying Poly/ML compiler / runtime
system have managed to follow the multicore trend. Isabelle2008 (June 2008)
was the first official release to support parallel proof processing in
\emph{batch mode} by default. At the same time it became apparent that
user-interfaces for parallel proof assistants require significant reworking
of the interaction model.

\item In 2010 the new Isabelle/PIDE concepts, with the underlying
Isabelle/Scala infrastructure, and Isabelle/jEdit as experimental
application, were presented in public at UITP 2010 \cite{Wenzel:2010}. That
was 2 years after the first concrete ideas for it had emerged, but still
more than 1 year to go before the first ``stable release'' of Isabelle/jEdit
with Isabelle2011-1 (October 2011).

\item In 2012 that initial release of Isabelle/jEdit was presented as system
description and tool demonstration at CICM \cite{Wenzel:2012:CICM} and some
of its concepts were explained at the co-located UITP 2012
\cite{Wenzel:2012:UITP-EPTCS}.

\item The subsequent releases of Isabelle/jEdit in May 2012, February 2013,
and November/December 2013 have consolidated the PIDE concepts and its
implementation. So many new things were introduced each time, that users
have occasionally complained about having to re-learn the Prover IDE with
each Isabelle release.

\end{itemize}

\medskip The current official release Isabelle2014 (August 2014) is
available from \url{http://isabelle.in.tum.de/website-Isabelle2014} for
Linux, Windows, Mac OS X. This paper is dedicated to some of its
newly-introduced PIDE concepts; the extended and updated Isabelle/jEdit
manual \cite{isabelle-jedit-manual} provides further information for
end-users. Isabelle/jEdit in Isabelle2014 also includes a new Simplifier
Trace panel with an interactive view of the simplification process; this was
contributed by Lars Hupel, see \cite{Hupel:2014}.

The Isabelle2014 distribution is notable as Isabelle/jEdit is now the sole
user interface by default: it is the main \verb|Isabelle2014|
application (specifically for each operating system family). The \verb|isabelle jedit| command-line tool may be used as well. The old \verb|isabelle emacs| tool for Proof General is \emph{not} included anymore, but
it is still available as optional component that needs to be downloaded
separately and configured manually. The \verb|isabelle tty| tool for
raw READ-EVAL-PRINT access to Isabelle/Isar on the terminal has been
discontinued altogether.%
\end{isamarkuptext}%
\isamarkuptrue%
\isamarkupsection{Asynchronous print functions \label{sec:async-print}%
}
\isamarkuptrue%
\begin{isamarkuptext}%
Asynchronous print functions in the PIDE document model were already
introduced in the release of Isabelle2013-2 (December 2013), and refined
further for Isabelle2014 (August 2014). The concept combines \emph{user
interaction} and \emph{tool integration} as explained for Isabelle2013-2 in
\cite{Wenzel:2014:ITP-PIDE}.

The general approach is to continue the reforms of READ-EVAL-PRINT
\cite{Wenzel:2012:UITP-EPTCS} as follows \cite[\S5]{Wenzel:2014:ITP-PIDE}:

\begin{itemize}

\item Edits may add or remove PRINT operations, without disturbing the
corresponding EVAL tasks. This principle of \emph{monotonicity} avoids
interruption of tasks that are still active in the document model, and
allows to use long-running or potentially non-terminating tools as print
functions. Typically these are automated provers (via Sledgehammer) or
disprovers (Quickcheck, Nitpick).

\item Activation or deactivation of PRINT tasks is subject to the
\emph{document perspective}. The whole theory library that is edited might
be big, but only small parts are visible in the editor. PIDE document
processing takes the open text windows as indication where to invest
resources for continuous processing. Various declarative parameters control
print functions that are implemented in user-space of Isabelle/ML: startup
delay, time limit, task priority, persistence of results within the document
model.

\item Support for explicit \emph{document overlays}, which are print
functions with arguments provided by some GUI components. This recovers the
appearance of direct access to command execution in the prover, despite the
thick layers of asynchronous PIDE protocol between the stateless/timeless
prover and the physical editor.

\end{itemize}

The screenshots \figref{fig:auto-tools} and \figref{fig:sledgehammer}
illustrate the use of asynchronous print functions and document overlays in
practice.

\medskip \Figref{fig:auto-tools} shows the result of Quickcheck, as an
example for \emph{automatically tried tools} that operate on outermost goal
statements (e.g.\ \hyperlink{command.lemma}{\mbox{\isa{\isacommand{lemma}}}}, \hyperlink{command.theorem}{\mbox{\isa{\isacommand{theorem}}}}), independently of
the state of the current proof attempt. Such tools work implicitly without
arguments, but there are global options in \emph{Plugin Options / Isabelle /
General / Automatically tried tools}. Results are output as
\emph{information messages}, which are indicated in the text area by blue
squiggles and a blue information sign in the gutter of the text window. The
message content may be shown as for other prover output in a separate
window. Some tools produce output with \emph{sendback} markup, which means
that clicking on certain parts of the message inserts that into the source
in the proper place.

\begin{figure}[htb]
\begin{center}
\includegraphics[scale=0.333]{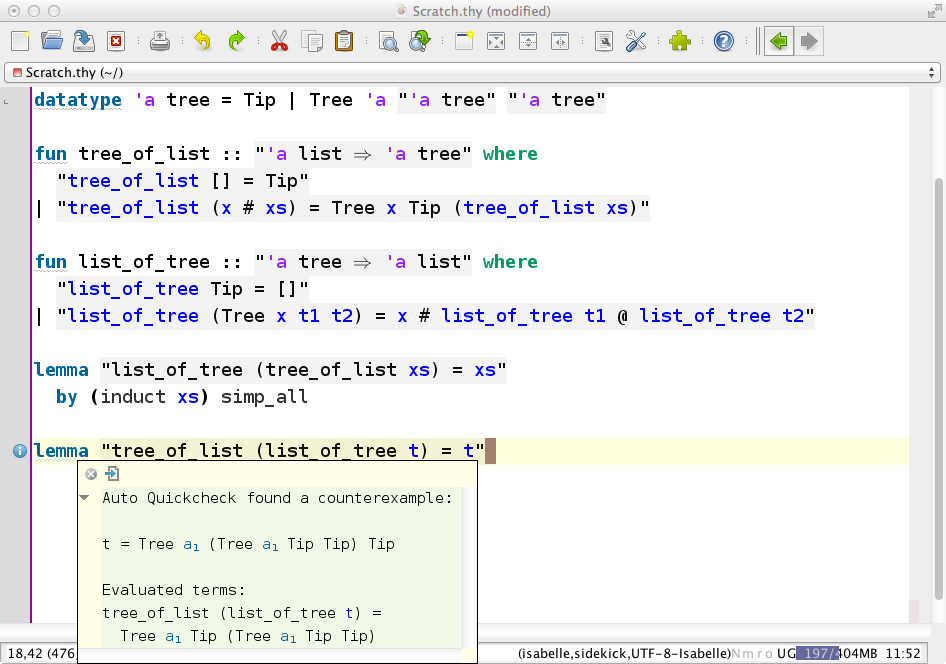}
\end{center}
\caption{Quickcheck as automatically tried tool}
\label{fig:auto-tools}
\end{figure}

\medskip \Figref{fig:sledgehammer} shows the \emph{Sledgehammer} panel,
which provides a view on some independent execution of the Isar command
\hyperlink{command.sledgehammer}{\mbox{\isa{\isacommand{sledgehammer}}}}, with process indicator (spinning wheel) and GUI
elements for important Sledgehammer arguments and options. Any number of
Sledgehammer panels may be active, according to the standard policies of
jEdit window management. Closing such a dockable window also cancels the
corresponding prover tasks.

\begin{figure}[htb]
\begin{center}
\includegraphics[scale=0.333]{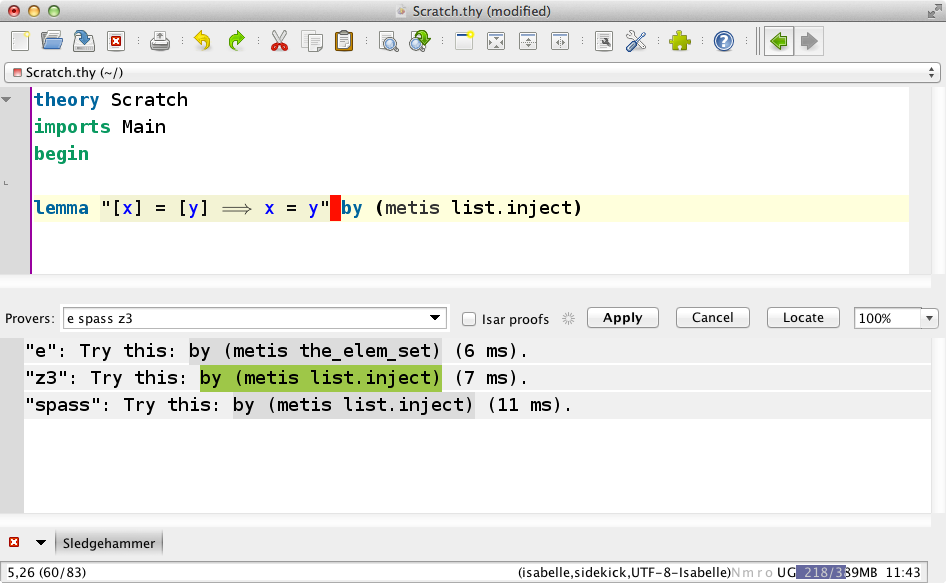}
\end{center}
\caption{An instance of the Sledgehammer panel}
\label{fig:sledgehammer}
\end{figure}

Technically, the Sledgehammer panel is a conventional GUI component on the
surface, but it is connected to the PIDE document model by producing some
document overlay when the user pushes the \emph{Apply} button. This leads to
some document edit that attaches a suitable asynchronous print function
(with arguments taken from the GUI panel), and forks some print task on the
prover side. Any output from that task is incrementally shown in the GUI
panel. The \emph{Cancel} button uses the execution id of the running print
operation to interrupt it on demand.

The overall interaction of the PIDE front-end with the prover back-end does
not prevent the user from editing the text nor the prover from checking
proofs in parallel. The only impact is some loss of performance to other
tools in the background, but this can be balanced via global system options
to adjust to the available number of cores.%
\end{isamarkuptext}%
\isamarkuptrue%
\isamarkupsection{Syntactic and semantic completion \label{sec:completion}%
}
\isamarkuptrue%
\begin{isamarkuptext}%
Semantic completion, based on authentic information from the proof
context, has been a ``nice to have'' features over several years. It was not
immediately obvious to teach that trick to a traditional LCF-style proof
assistant like Isabelle, which was not made for that 25 years ago.

Even just the editor GUI part of auto completion has turned out much less
trivial than anticipated in 2009/2010 \cite{Wenzel:2010}, where the (naive)
idea was to connect to an existing completion plugin of jEdit. Over the last
5 years the completion mechanism in Isabelle/jEdit has changed several
times, but various problems with the timing of GUI events still occur in
Isabelle2013-2.

Completion intercepts the regular key event handling of the main text area,
and needs to work smoothly as the user is typing slowly or quickly. The
completion popup changes the keyboard focus to a different component, which
can lead to odd effects of loosing key events in a situation where the user
is typing fast, but the graphics display is too slow to catch-up (e.g.\ due
to bad X11 rendering performance, which can happen both for local and remote
displays).

The lesson learned here is that a Prover IDE is a highly interactive
computer-game, with demands of real-time reactivity that were not present in
TTY front-ends from the past.

\medskip Both the GUI event handling and the semantic aspects of completion
in Isabelle/jEdit have been significantly reworked for Isabelle2014,
according to the following general principles.

\begin{itemize}

\item \textbf{Syntactic completion} is based on information that is
\emph{immediately} available in the editor, e.g.\ keyword tables for certain
sub-languages of Isabelle, like the so-called ``outer syntax'' of
Isabelle/Isar, or Isabelle/ML. Completion for Isabelle symbols is an
important a special case of this: when the user types ``\verb|==>|''
he normally expects to get ``\isa{{\isasymLongrightarrow}}'' within formal text.

\item \textbf{Semantic completion} is produced by the prover
\emph{eventually}, after a full round-trip through the asynchronous PIDE
protocol. This information usually arrives with a delay of 100--500\,ms and
is then merged with the available syntactic completion, before it is used
for GUI rendering (e.g.\ for emphasis of text or a popup).

\item \textbf{Completion markup} may be produced by the prover in any of the
following forms:

\begin{itemize}

\item \textbf{Language context} guides the syntactic completion. Isabelle is
a framework of many sub-languages, which have different requirements for
completion. The language context for some text range informs the editor
about the language name (e.g.\ to use a different keyword table), and some
common flags like use of Isabelle symbols and antiquotations.

For example, the term language in Isabelle supports symbols, but no
antiquotations. In contrast, the document language (a semi-formal version of
{\LaTeX}) supports antiquotations, but no symbols. An antiquotation that
puts a term inside some document source needs to switch the language context
accordingly, and several such changes of language context can happen in a
small piece of theory source.

This approach already works smoothly for text that is structurally mostly
correct, but a special challenge of PIDE interaction is to treat situations
of partial or broken input gracefully. The expectation of the user versus
the system may disagree about the intended structure of some unfinished
text.

\item \textbf{Completion items} result from failed name space lookups of
formal entities (type names, term constants, fact names etc). Luckily the
prover already has a mostly uniform concept of name spaces, in order to
intern names given by the user to the actual formal entities from the
context. The error situation has been slightly modified to include a list of
alternative names into the error message, as PIDE markup that is not
immediately visible, but available to the completion mechanism.

For performance reasons, it is important to produce completion items only
for failed name-space lookups, which are relatively rare, and not for the
majority of successful ones. There is nonetheless a simple way for the user
to request more information: adding a suffix of underscores to a partial
name provokes an error with extra completion information. A double
underscore on its own serves as wildcard to query the whole name space, but
output is always truncated to a reasonable limit for display. Explicit
completion requests via underscores are particularly important for the term
language, because undeclared constants alone are accepted as free variables,
without any error nor completion information.

\item \textbf{No-completion zones} enable the prover to \emph{negate}
already discovered syntactic completions of the editor. Such non-monotonic
change of the meaning of incremental document content is always critical,
and can lead to erratic behaviour. Here it should be seen as a feature of
last resort, to suppress odd effects when Isar keywords like ``\verb|:|'' and ``\verb||\verb,|,\verb||'' should remain like that, and not be offered
as completion candidates for symbols ``\isa{{\isasymin}}'' and ``\isa{{\isasymor}}''.

\end{itemize}

\end{itemize}

\textbf{Spell-checking} is another application of the same PIDE
infrastructure, which is somewhere in between syntactic and semantic
completion. Based on prover markup for the language context, e.g.\ to
determine ranges of prose text inside document sources or comments, the
editor uses a conventional dictionary-based spell-checker to propose
alternatives to words spotted in the text. This is important to write books
and papers based on Isabelle theory sources, which is in fact the most
relevant practical application of Isabelle over 15 years.

\Figref{fig:completion} illustrates spell-checking within informal text: the
default dictionary does not know about \verb,Hilbert's,, but this is not an
error, merely highlighting. Moreover there is semantic completion within the
term language, using an extra underscore to let the prover expose
constants from the theory context.

\begin{figure}[htb]
\begin{center}
\includegraphics[scale=0.333]{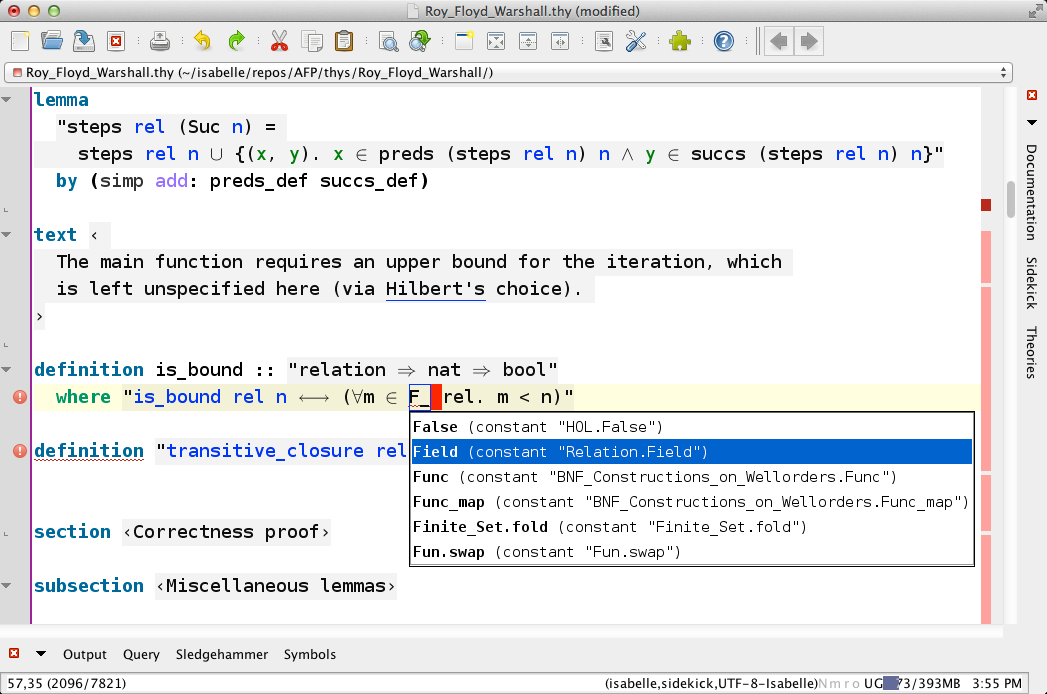}
\end{center}
\caption{Spell-checking within informal text and semantic completion
  within terms}
\label{fig:completion}
\end{figure}%
\end{isamarkuptext}%
\isamarkuptrue%
\isamarkupsection{Editor navigation \label{sec:navigation}%
}
\isamarkuptrue%
\begin{isamarkuptext}%
PIDE document content consists of sources that are augmented by
semantic markup from the prover, as explained in \cite{Wenzel:2011:CICM}.
The abstract syntax for the markup follows untyped XML, and the semantics is
often close to hypertext, with occurrences of formal entities in defining or
referencing positions. Thus it is rather obvious to think of standard
XML/HTML rendering and browsing of PIDE documents.

In fact, early versions of Isabelle/jEdit (from 2010 to 2012) were using a
basic HTML4 rendering engine, always with the anticipation for the HTML5
browser component that was promised by Sun, and delivered at last by Oracle
for Java 7. None of this is used in Isabelle/jEdit today, because it
introduces more problems than it solves: HTML is a very complex collection
of standards in many versions and different implementations. Professional
Web designers (and their tools) know how to cope with major browsers, but
exotic HTML components for Java/Swing or JavaFX can hardly be expected to
achieve professional quality.

Already since 2013, Isabelle/jEdit uses a slightly augmented version of the
main jEdit text area, with specific support for \emph{active areas}.
Hyperlinks are an important special case of that: prover markup is turned
directly into familiar clickable spots in the text (via mouse hovering with
the CONTROL or COMMAND modifier key pressed). In 2014 the visual appearance
approximates that of major Web browsers further, e.g.\ due to change of the
mouse pointer. There is now also the long-missing connection to an existing
\emph{Navigator} plugin from the jEdit repository, which is a rare case of
successful re-use of software components: no special tricks nor
reconsideration of the underlying concepts were required, to make
Isabelle/jEdit converge with regular HTML browsers in this respect.

\begin{figure}[htb]
\begin{center}
\includegraphics[scale=0.333,valign=t]{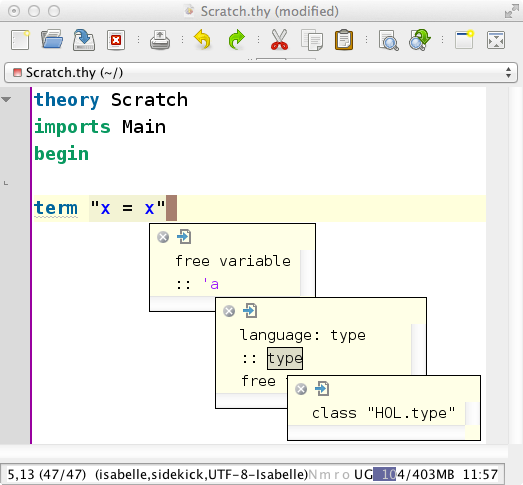}
\includegraphics[scale=0.333,valign=t]{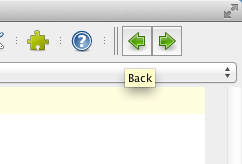}
\end{center}
\caption{Browsing PIDE document content via tooltips and hyperlinks}
\label{fig:navigation}
\end{figure}

\medskip Isabelle/jEdit text areas with markup and hyperlinks are used
uniformly wherever that makes sense: for the main editor buffer, output
panels, tooltips etc. The user who sees a printed term somewhere can follow
the implicit links to the definitions of the formal entities shown there,
and return easily to the original editor location via the now standard
\emph{Back} button in the toolbar, see \figref{fig:navigation}.%
\end{isamarkuptext}%
\isamarkuptrue%
\isamarkupsection{Auxiliary files within the document-model \label{sec:aux-files}%
}
\isamarkuptrue%
\begin{isamarkuptext}%
Ultimately, the main job of an IDE is to manage a collection of
sources and the results of processing them seamlessly, taking implicit and
explicit structural dependencies into account. So far the PIDE document
model was based on two levels in the structural hierarchy: an acyclic graph
of \emph{document nodes} (theories), where each node consists of a list of
\emph{command spans} (like in Proof General \cite{Aspinall:TACAS:2000}).

\begin{figure}[!htbp]
\begin{center}
\includegraphics[scale=0.333]{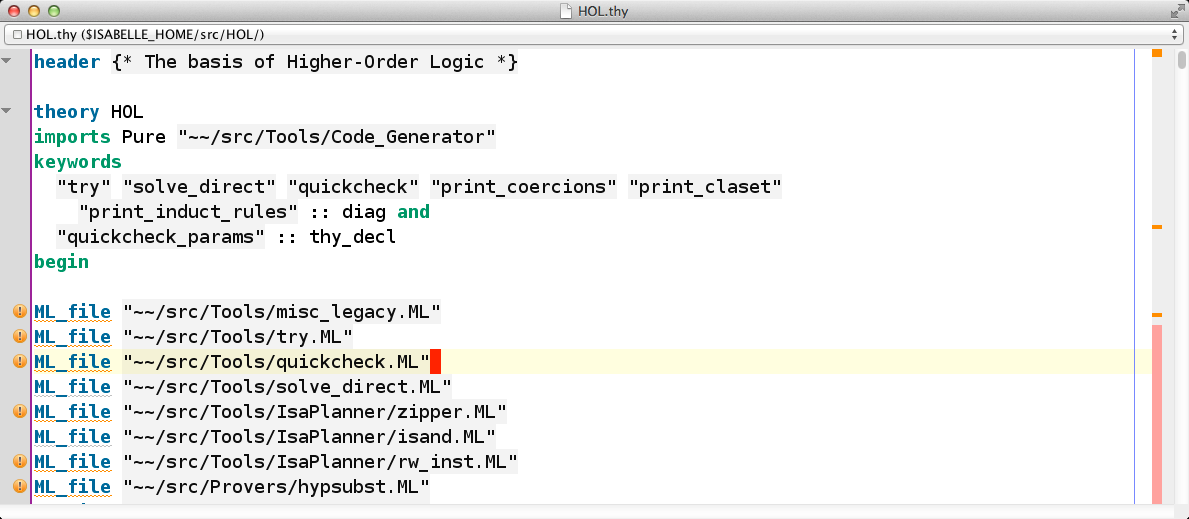}

\vspace*{1ex}

\includegraphics[scale=0.333]{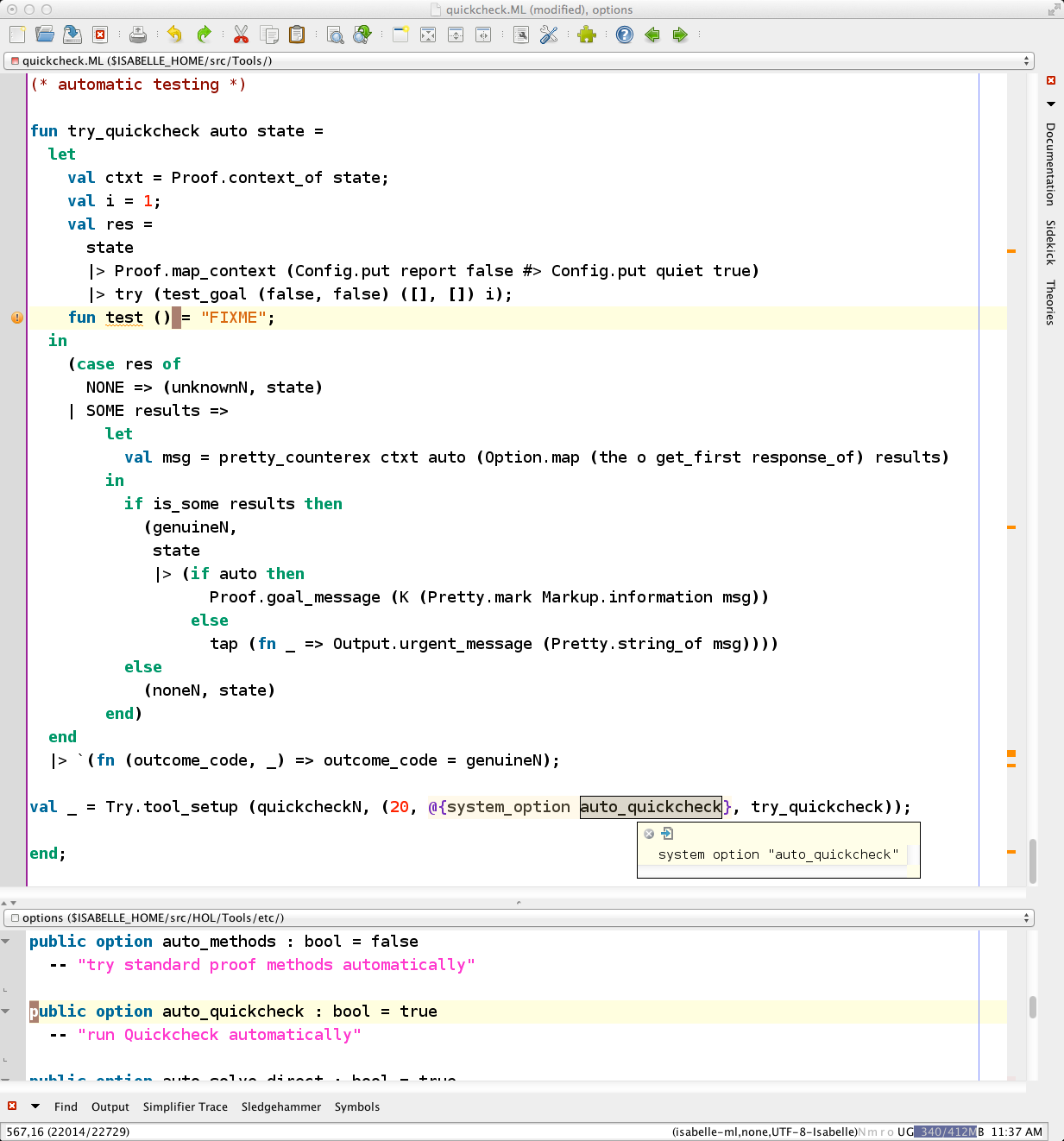}
\end{center}
\caption{Live editing and browsing of Isabelle/HOL ML files}
\label{fig:ml}
\end{figure}

Apart from that, it was always possible to refer to auxiliary files as a
semi-official feature addition, but with limited management in the IDE. That
unsatisfactory situation has ended, and there is now first-class support for
auxiliary files that appear as arguments to special \emph{load commands}
inside document nodes. Thus the source text is conceptually extended by
so-called \emph{text chunks} that are stored elsewhere, and may be edited /
loaded / saved independently of the theory itself. The Prover IDE takes care
to forward the correct version of auxiliary file content to the prover as a
\emph{blob}, but without using the global file-system.\footnote{By-passing
the file-system is an important PIDE principle to avoid statefulness and
restricting the document-model to a single version. Here the jEdit buffer
management takes over this role: the current editor content is propagated to
the prover as latest version, while further changes may follow.}

\medskip This extra file management is particularly relevant for development
of Isabelle/HOL itself within the Prover IDE. According to usual practice of
LCF-style proof assistants, the main logical environment emerges by
alternating theory specifications with ML modules. It is now possible to use
Isabelle/jEdit to explore the inner workings of Isabelle/HOL modules and
their dependencies on theory content, notably in conjunction with the
navigation support explained above (\secref{sec:navigation}).

For live editing of the Isabelle/HOL sources, the logic session image of
Isabelle/jEdit needs to be set to \verb|Pure|, which requires a restart
of the Prover IDE. Then any of the theories may be opened, e.g.\ \verb|$ISABELLE_HOME|\verb|/|\discretionary{}{}{}\verb|src|\verb|/|\discretionary{}{}{}\verb|HOL|\verb|/|\discretionary{}{}{}\verb|HOL.thy| by using that path notation literally in
the jEdit file browser (even on MS Windows). The \hyperlink{command.ML-file}{\mbox{\isa{\isacommand{ML{\isacharunderscore}file}}}} commands
in such theories refer to Isabelle/ML modules that are compiled on the spot.
By default the prover reads the source from the file-system, but by
following the implicit hyperlink of the file argument (or opening files in
the jEdit file-browser) the editor takes over the responsibility for the
sources and its subsequent changes.

Thus the user may edit Isabelle/ML source files, without ever saving the
content, while the Poly/ML compiler provides continuous feedback on
warnings, errors, name references, inferred types etc.\ as part of the PIDE
document model (see \figref{fig:ml}). This works reasonably well for source
files up to 100\,KB each. The total volume of ML sources contributing to
Isabelle/HOL is actually so high that its cumulative PIDE markup requires
more than 2 GB Java heap space. This performance bottle-neck is addressed by
some special tricks with asynchronous print functions
(\secref{sec:async-print}) which were introduced in 2013 for quite different
applications. Here the mechanism is re-used as follows: Poly/ML compiler
markup is stored in compact form within the ML process, and only reported to
the editor when the corresponding ML file becomes part of the visible
perspective. The document markup is removed from the editor process when
visibility gets lost.

Thus the massive amount of PIDE markup produced by the ML compiler is
``swapped-in'' and ``swapped-out'' on demand, without changing the content
of the ML environment itself. Consequently the full Isabelle/HOL bootstrap
environment can be edited with full Poly/ML markup, even on small computers
with only 4--8 GB memory.

\medskip As a corollary to this scalable approach to continuous editing and
compilation of Isabelle/ML files, there is also support for official
Standard ML via the \hyperlink{command.SML-file}{\mbox{\isa{\isacommand{SML{\isacharunderscore}file}}}} command. Thus Isabelle/jEdit can be
used as IDE for SML'97, without any connection to theory or proof
development. The two ML environments are managed independently within the
same runtime system, but there are also simple means to exchange toplevel ML
bindings, e.g.\ to re-use the parallel functional programming library of
Isabelle/ML in Standard ML, or to print messages in Standard ML that are
recognized by the Prover IDE for its \emph{Output} panel. The Prover IDE
provides some simple examples for that in its \emph{Documentation} panel, in
the entry for \verb|$ISABELLE_HOME|\verb|/|\discretionary{}{}{}\verb|src|\verb|/|\discretionary{}{}{}\verb|Tools|\verb|/|\discretionary{}{}{}\verb|SML|\verb|/|\discretionary{}{}{}\verb|Examples.thy| that is
only a single click away from direct editing and browsing. Here the
enclosing theory is used like a project file for SML modules, with the
possibility to add extra explanations around it.

Thus Standard ML has gained a reasonably modern IDE after some decades of
waiting. Navigation of the sources works as usual (\secref{sec:navigation}),
but semantic completion (\secref{sec:completion}) is still missing because
Poly/ML (version 5.5.2) does not provide that information. Also lacking is
support for the interactive debugger of Poly/ML. Such further fine-grained
interaction with the ongoing execution process would be quite useful for
other Isabelle languages as well, e.g.\ to analyse the behaviour of tactical
expressions beyond the single-stepping of outermost commands or adhoc
printing of intermediate states.

\medskip If that train of thought is continued further, it could meet with
recent trends towards ``live programming'', as advocated by Bret Victor on
his blog\footnote{\url{http://worrydream.com}}, for example. This is a
revival and continuation of older ideas from Smalltalk of the 1970s and
1980s, but adapted to the possibilities of the hardware from today.
Interactive theorem proving has always been conceptually close to such
``dynamic'' development models, and the removal of the TTY loop in
Isabelle/PIDE could help to demonstrate that in reality.

A major difference to Smalltalk and classic object-oriented programming is
that Isabelle/PIDE document content is immutable and processed
monotonically, in a timeless and stateless manner, with functional update of
pure data. That used to be costly in the past, but on the multicore hardware
of today immutability is a big asset.%
\end{isamarkuptext}%
\isamarkuptrue%
\isamarkupsection{Conclusion%
}
\isamarkuptrue%
\begin{isamarkuptext}%
This paper has covered the main novelties of Isabelle/PIDE and its
Isabelle/jEdit application from the past 2 years. The total lifespan of the
project so far has been approx.\ 6 years. This relatively late stage is
characterized by cumulative and incremental improvements, towards a more and
more consolidated prover interface that users find hard to avoid.

With the ``self-application'' of the IDE to develop ML modules of Isabelle
itself, the name of \emph{integrated development environment} is really
justified. Ambitious users may even invent their own add-on languages within
PIDE documents, and apply the same principles of continuous editing and
processing with rich markup. This can be done with or without connection to
the formal content of Isabelle theories, as demonstrated by the IDE for
Standard ML that is now included in Isabelle2014.

\medskip Despite the growing size of the system, it is unlikely that it will
ever get finished: the more it advances, the more requests from users to add
features. Some important aspects for the future are as follows:

\begin{itemize}

\item Full integration of the Isabelle document preparation system into the
IDE.

Presently the document sources are edited in Isabelle/jEdit as {\LaTeX}
chunks within theory files, while the build process works in batch-mode.
This used to be a system shell invocation of \verb|isabelle build|,
but now also works within the jEdit Console plugin via \verb|Build.build| in Scala. At the next stage it should be instantaneous within
the running PIDE session, without demanding an extra build job.

\item Support for a remote prover process via SSH socket connection.

Veterans of early Proof General may remember the old RSH facility (long
before SSH came into existence), which was important to work on big
projects. The prover back-end would run remotely on a server-class machine,
and the editor front-end locally on a normal workstation (or laptop). This
traditional distribution of the workload becomes important again, as the
consumer market stagnates at relatively small number of cores (4--8, maybe
16).

There is also a pending problem of memory size: present mid-range machines
are equipped with 8\,GB, sometimes 16\,GB. This works for medium and big
applications, where the prover is still running tightly in 32\,bit mode and
3--4\,GB maximum memory. Really large Isabelle applications already require
64\,bit mode, and thus need double memory size before any benefit can
happen. The memory footprint of the Isabelle/jEdit alone is 4--8\,GB. This
means that 32\,GB or more is required, but it is often seen on remote
servers only.

\item Genuine support for structured proof editing within the IDE.

So many conceptual problems and technical side-conditions had to be
addressed over the past 6 years of PIDE that proper support for proof
editing has been neglected. The Isabelle/Isar proof language with its rich
structure is predestined to go beyond mere ``proof scripts'', but it still
needs to be done seriously. For example, there should be \emph{smart
templates} (e.g.\ for induction proofs), depending on the present document
source within the editor and its partial processing by the prover.

\end{itemize}

The near future should also see the final disposal the TTY loop, which
merely serves as optional legacy feature in Isabelle2014, and needlessly
complicates the implementation.%
\end{isamarkuptext}%
\isamarkuptrue%
\isadelimtheory
\endisadelimtheory
\isatagtheory
\isacommand{end}\isamarkupfalse%
\endisatagtheory
{\isafoldtheory}%
\isadelimtheory
\endisadelimtheory
\isanewline
\end{isabellebody}%

\bibliographystyle{eptcs}
\bibliography{root}

\end{document}